\documentclass[aps,prb,twocolumn,superscriptaddress,preprintnumbers,amsmath,amssymb]{revtex4-2}
\usepackage{graphicx}
\usepackage{dcolumn}
\usepackage{bm}
\usepackage{subfigure}
\usepackage{epstopdf}
\usepackage[hyperfigures=true,bookmarksnumbered=true,bookmarksopen=true,bookmarks=true,linkcolor=blue,urlcolor=blue,citecolor=blue,colorlinks=true,pdfhighlight=/O,pdfstartview={XYZ null null 1}]{hyperref}

\usepackage[colorlinks=true, linkcolor=blue, citecolor=blue, urlcolor=blue]{hyperref}

\begin{document}


\title{Tuning Magnetic and Electronic Properties of Double Perovskite La$_2$CoIr$_{1-x}$Ti$_x$O$_6$}

\author{Sromona Nandi}
\altaffiliation{These authors contributed equally to this work.}
\affiliation{Department of Physics, Indian Institute of Technology Tirupati, Tirupati, Andhra Pradesh 517619, India}

\author{Vineeta Yadav}
\altaffiliation{These authors contributed equally to this work.}
\affiliation{Department of Physics, Indian Institute of Technology Tirupati, Tirupati, Andhra Pradesh 517619, India}

\author{Sheetal}
\affiliation{School of Physical Sciences, Indian Institute of Technology Mandi, Kamand, Mandi, Himachal Pradesh 175075, India}

\author{C. S. Yadav}
\affiliation{School of Physical Sciences, Indian Institute of Technology Mandi, Kamand, Mandi, Himachal Pradesh 175075, India}

\author{Bikash Das}
\affiliation{School of Physical Sciences, Indian Association for the Cultivation of Science, 2A \& 2B Raja S. C. Mullick Road, Jadavpur, Kolkata 700032, India}

\author{Subhadeep Datta}
\affiliation{School of Physical Sciences, Indian Association for the Cultivation of Science, 2A \& 2B Raja S. C. Mullick Road, Jadavpur, Kolkata 700032, India}

\author{Kapildeb Dolui}
\affiliation{Department of Physics, Indian Institute of Technology Tirupati, Tirupati, Andhra Pradesh 517619, India}

\author{Rudra Sekhar Manna}
\email[]{rudra.manna@iittp.ac.in}
\affiliation{Department of Physics, Indian Institute of Technology Tirupati, Tirupati, Andhra Pradesh 517619, India}

\date{\today}

\begin{abstract}

The La$_2$CoIr$_{1-x}$Ti$_x$O$_6$ ($0 \leq x \leq 1$) double perovskite series serves as an effective platform for investigating the evolution of magnetic and electronic properties as a function of chemical pressure (doping) or hydrostatic pressure due to the interplay between the electrons correlation ($U$) and spin–orbit coupling (SOC). In this study, the substitution of nonmagnetic Ti$^{4+}$ at the magnetic Ir$^{4+}$-site leads to a systematic decrease in unit cell volume keeping the monoclinic $P2_1/n$ symmetry throughout, reflecting the effect of chemical pressure along with a gradual suppression of magnetic interactions. The parent compound ($x = 0$) exhibits a ferromagnetic-like state with a Curie temperature, $T_\mathrm{C}$ = 92 K, which continuously evolves into an antiferromagnetic ground state upon full Ti substitution ($x = 1$) with a N\'eel temperature, $T_\mathrm{N}$ = 14.6 K. Isothermal magnetization measurements reveal a hysteresis behavior with step-like feature at zero field, indicative of a noncollinear magnetic ordering. Additionally, the enhancement of magnetization under hydrostatic pressure on La$_2$CoIrO$_6$ ($x = 0$) suggests the presence of piezomagnetic behavior. Thermal expansion measurements on La$_2$CoIrO$_6$ highlight a coupling between spin and lattice degrees of freedom. The pressure dependence of the transition temperature in the zero-pressure limit, calculated using Ehrenfest’s relation, shows good agreement with magnetization data under applied pressure. First-principles density functional theory (DFT) calculations preformed for $x$ = 0, 0.5 and 1, further reveal that strong SOC associated with Ir plays a decisive role in shaping the electronic band structure, with the insulating gap progressively widening as Ti content increases from 0.28 eV ($x$ = 0), 0.44 eV ($x$ = 0.5), and 1.01 eV ($x$ = 1). The magnetic moment decreased more than 50\% for $x$ = 0.5, showing the decrease in magnetic exchange pathways. Collectively, these results establish La$_2$CoIr$_{1-x}$Ti$_x$O$_6$ as a model system for exploring the interplay among electrons' correlation, SOC, chemical pressure, and doping-driven magnetic phase transitions, providing valuable insights into the tunability of electronic and magnetic properties in complex oxides.

\end{abstract}

\maketitle

\section{Introduction}

Double perovskite oxides, denoted as Ln$_2$MM$^\prime$O$_6$, where Ln represents rare earth elements or alkaline earth metals, M and M$^\prime$ be the transition metal elements, are of interest due to their cationic ordering, antiphase boundaries, and multiple exchange interactions in addition to their intriguing physical properties such as higher Curie-temperatures, metal-insulator transitions, different magnetic orderings, and structural and magnetic phase transitions~\cite{Kato2002,KatoPRB2002}. In recent years, more attention has been diverted to the iridium (Ir) based oxides, which have extended 5$d$ orbitals leading to strong spin-orbit coupling (SOC) that are comparable to onsite Coulomb repulsion ($U$) and crystal electric field (CEF) interactions. Following the discovery of a spin-orbital Mott insulating state in the layered iridate Sr$_2$IrO$_4$, a number of Ir$^{4+}$ (5$d$) iridates with a $J_\mathrm{{eff}}$ = 1/2 ground state~\cite{Kim2008} demonstrate that the delicate balance between interactions causes exotic magnetic and dielectric behavior~\cite{Kim2009,Cook2015,Aczel2016,Jackeli2009,Watanabe2013}. 

An iridium-based double perovskite La$_2$CoIrO$_6$ (LCIO hereafter) where the magnetism comes from both Co$^{2+}$ and Ir$^{4+}$, shows a ferromagnet-like order below $\sim$90 K~\cite{Narayanan2010,Lee2015} with a band gap of 0.26 eV, driven by strong Co–Ir sublattice coupling and a step-like feature in isothermal magnetization hinted noncollinear spin structures~\cite{Narayanan2010}. Frustration driven short-range magnetic correlation much above magnetic transition by means of NMR and ESR measurements~\cite{Iakovleva2018} and a reentrant spin-glass behavior below the ordering temperature via AC susceptibility measurements have been reported~\cite{Song2017}. There are attempts to tune the magnetism and transport properties by doping Sr~\cite{Narayanan2010} and Ca~\cite{Courtim2015} at the A-site (La in this case) due to the interplay between structural distortion, spin–orbit coupling, and cation disorder. A composition- and temperature-driven structural transition from $P2_1/n$ $\leftrightarrow$ $P2_1/n$ + $I2/m$ $\leftrightarrow$ $I2/m$ $\leftrightarrow$ $ I4/m$ $\leftrightarrow$ $Fm\bar{3}m$ remarked in Sr-doped La$_{2-x}$Sr$_x$CoIrO$_6$ ascribed to an increase in average size of A-site and a band gap reduction from 0.26 eV to 0.05 eV attributed to hole doping and broadening of bandwidths. There is a decrease in the ordering temperature from 90 K ($x = 0$) to 70 K ($x = 2$). In addition, the ferromagnetic component is reduced two orders of magnitude with doping which favors collinear spin alignment~\cite{Narayanan2010}. Conversely, Ca doping, La$_{2-x}$Ca$_x$CoIrO$_6$ decreases the ordering temperature  attributed to Co$^{2+}$ coexistence with Co$^{3+}$ ~\cite{Courtim2015}. In comparison, La$_2$CoTiO$_6$ (LCTO hereafter), an antiferromagnetic insulator with $T_\textrm{N}$ = 14.6 K in which the magnetic moment comes only from Co$^{2+}$ as Ti has 4+ oxidation state. By the application of hydrostatic pressure, the insulating band gap of 1.01 eV at ambient pressure closes leading to a transition into an antiferromagnetic metallic state at 42 GPa which is accompanied by a spin-state transition of Co$^{2+}$ from it's high-spin to low-spin state. With further pressure increase, the system reaches to a nonmagnetic state with complete quenching of the Co moment at 130 GPa~\cite{Nandi2024}. 

In this study, we probe the role of SOC through substitution of nonmagnetic Ti$^{4+}$ at the Ir-site, $i.e.$, La$_2$CoIr$_{1-x}$Ti$_x$O$_6$ using combined experimental and computational approaches. The substitution of Ti offers not only effect of SOC, but introduce chemical pressure to these 3$d$–5$d$ double perovskites. Since Ti$^{4+}$ is nonmagnetic and lacks SOC, increasing its concentration effectively dilutes the SOC-driven interactions originating from Ir$^{4+}$. This allows a systematic investigation of how weakening SOC influences the magnetic ordering, exchange pathways, and transport behavior. Such a study can provide valuable insights into the interplay between SOC, electronic correlations, and structural distortions in complex oxides. 

We prepared both the parent compounds (LCIO and LCTO) and the doping series La$_2$CoIr$_{1-x}$Ti$_x$O$_6$ for various $x$ values via the solid-state reaction method, followed by their structural and magnetization characterizations. X-ray diffraction (XRD) confirms that the series crystallizes into a monoclinic structure with a space group $P2_1/n$ (No.~14). As Ti content increases, the magnetic ordering temperature ($T_\textrm{C}$) systematically decreases from 92 K ($x = 0$, ferromagnetic-like) to 14.6 K ($x = 1$, antiferromagnetic). A step-like behavior at zero field is observed throughout the series with partial Ti content (except for $x = 0$) exhibited from the isothermal magnetization measurements, which is indicative of noncollinear magnetic ordering also disclosed in literature~\cite{Narayanan2010}. Thermal expansion measurements were performed on LCIO which reveal a pronounced, broad peak indicating a strong coupling between magnetic and lattice degrees of freedom. The equal area entropy construction is employed on the thermal expansion and heat capacity data~\cite{Narayanan2010_thesis} to estimate initial pressure dependence of the transition temperature which is in good agreement with the magnetization under pressure data. Magnetization under hydrostatic pressure measurements were conducted on LCIO for pressures up to $P$ = 2.22 GPa. There is a linear increase in the magnetic transition to 100 K ($P$ = 2.22 GPa) from 92 K ($P$ = 0 GPa), suggesting enhanced spin canting toward alignment with the external magnetic field as pressure increases. The piezomagnetic effect, described as the effect of pressure on the magnetization~\cite{Nanjo2025}, is observed in LCIO through the linear increase in magnetization with pressure. To investigate the influence of Ti doping on SOC and its impact on magnetic and electronic properties, Density Functional Theory (DFT) calculations have been performed for the parent compounds LCIO, LCTO, and the intermediate composition $x = 0.5$. Structural relaxations and energy minimization calculations on collinear magnetic configurations reveal that the system has an antiferromagnetic ground state, where Ir$^{4+}$ spins are coupled to Co$^{2+}$ spins. While LCIO exhibits metallic behavior in DFT (GGA), this discrepancy with experimental insulating (band gap 0.26 eV) behavior is resolved by incorporating $U$ and SOC. Ti doping leads to a reduction in Ir moments, an increased band gap (up to 0.44 eV for $x = 0.5$), and lattice volume contraction, consistent with experimental observations. Ti-$d$ states dominate the conduction band at higher doping levels, exceeding contributions from Co and Ir. Orbital-projected partial density of states (PDOS) further confirm the suppression of Ir-$d$ and Co-$d$ states near the Fermi level, indicating a transition toward an insulating state. These results highlight a strong interplay between composition, structure, and the electronic and magnetic properties. 

The remainder of this paper is organized as follows: In section~\ref{sectionII}, we present a detailed discussion of the crystal structures examined via X-ray diffraction (XRD) and analyzed using the FullProf software. The magnetic characterization of the La$_2$CoIr$_{1-x}$Ti$_x$O$_6$ series is described, along with magnetization measurements of LCIO under hydrostatic pressure and thermal expansion studies. Section~\ref{sectionIII} focuses on the transport and magnetic properties of the $x$ = 0, 0.5 and 1, as investigated within the framework of DFT. Finally, section~\ref{sectionIV} provides a summary of our findings and concludes the study.

\section{Experimental section} \label{sectionII}

\begin{figure*}[h!bt]
\centering
\includegraphics[width=2\columnwidth]{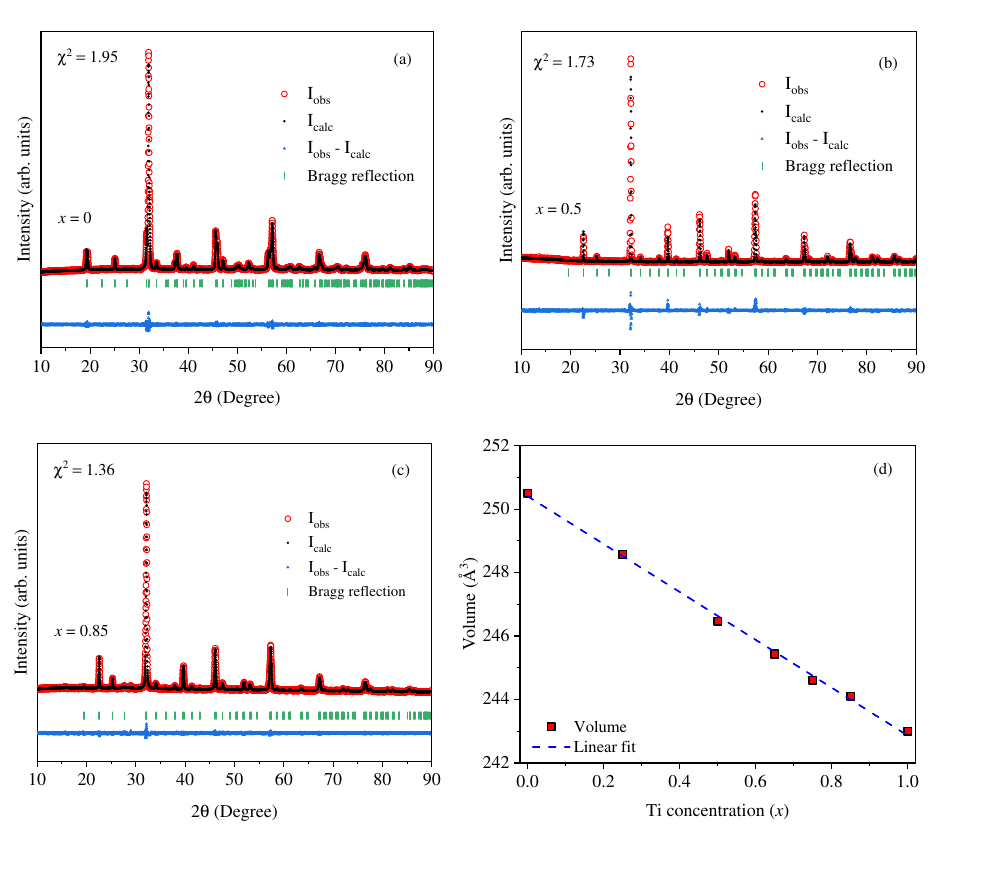}
\caption{\label{fig:xrd_plot.pdf} Rietveld-refined powder XRD patterns at room temperature for (a) $x = 0$, (b) $x = 0.5$, (c) $x = 0.85$, and (d) composition dependence of the unit cell volume of compounds fitted with linear Vegard's law. The black line, the red line, the vertical marks (green line), and the blue lines correspond to the experimental data, the calculated data, Bragg peaks, and the difference between the observed and calculated curves, respectively. The other concentrations ($x$ = 0.25, 0.65, and 0.75) are also synthesized in pure form and their lattice parameters are given in table~\ref{tab:table 1}. The detailed structural details of $x = 1$ has already been reported in our earlier paper~\cite{Nandi2024}.}
\end{figure*}

Polycrystalline samples of La$_2$CoIr$_{1-x}$Ti$_x$O$_6$ were synthesized by solid-state reaction for five different compositions along with the parent compounds with $x$ = 0, 0.25, 0.5, 0.65, 0.75, 0.85, and 1. Stoichiometric amounts of reactants La$_2$O$_3$ (Sigma, 99.99\%), TiO$_2$ (Sigma, 99.99\%), CoO (Sigma, 99.99\%), and IrO$_2$ (Sigma, 99.9\%) were mixed and heated at 1,050 $^{\circ}$C ($x = 0$) and 1,250 $^{\circ}$C ($x$ = 0.25, 0.5, 0.65, 0.75, 0.85, and 1) in air within a muffle furnace for 48 hours with one intermittent grinding. The materials were ground, pressed into pellets, and heated at 1,200 $^{\circ}$C ($x = 0$) and 1,300 $^{\circ}$C ($x$ = 0.25, 0.5, 0.65, 0.75, 0.85, and 1) for a week. Due to the hygroscopic nature of rare earth oxides, La$_2$O$_3$ was preheated at 900 $^{\circ}$C overnight to remove any possible moisture. X-ray powder diffraction data were collected using a Rigaku X-ray diffractometer using PANalytical Aeris XRD in IIT Tirupati with a 2$\theta$ step size of 0.01. Rietveld refinements of the structural parameters were performed using the FullProf suite software~\cite{Nandi2024}. Magnetization measurements were carried out using a Magnetic Property Measurement System (MPMS 3, Quantum Design) at IIT Mandi and a Physical Property Measurement System attached with Vibrating Sample Magnetometer (PPMS-VSM, DynaCool, Quantum Design) at IACS, Kolkata. Magnetization as a function of temperature (both in zero field cooled (ZFC) and field cooled (FC) modes) and isothermal magnetization below the magnetic ordering temperature were carried out for the entire doping series. Magnetization under hydrostatic pressure measurements were performed up to 2.22 GPa for LCIO in a CuBe pressure cell~\cite{Tateiwa2011,Tateiwa2013} placed inside a SQUID (MPMS, Quantum Design) magnetometer. All measurements were performed upon compression as it was not possible to control the pressure while decompression. Daphne 7373 oil was used as a pressure transmitting medium. One small piece of lead ($\sim$0.1 mg) was placed together with the sample inside the pressure cell and another piece ($\sim$0.1 mg) was placed outside the pressure cell. As the superconducting transition temperature of lead (kept inside the pressure cell) decreases with the increase of pressure, the difference in transition temperature with respect to the lead which is kept outside (at ambient pressure) helped us to determine the exact pressure inside the pressure cell. The gasket of the pressure cell contained the sample of mass $\sim$0.35 mg and a lead piece of mass $\sim$0.1 mg. The empty cell background data was subtracted~\cite{Tateiwa2011} by using the automatic background subtraction (ABS) procedure. Measurements of lead and sample were performed with the fields of 2 mT and 0.1 T respectively. Thermal expansion measurements were conducted using high-resolution capacitive dilatometer~\cite{Barron1980,Manna2012,Kuechler2012} which was integrated into the multi-function PPMS probe, allowing for the detection of length changes (${\Delta L(T)}$) as small as 0.05 $\mathrm{\AA}$ across a sample with a length ($L_{0}$) spanning several millimeters. 

\begin{table*}[h!tb]
\centering
\setlength{\tabcolsep}{10pt} 
\renewcommand{\arraystretch}{1} 

\begin{tabular}{c c c c c c c c c}
\hline
\hline
\multicolumn{1}{c}{Structural} & \multicolumn{7}{c}{Ti concentration ($x$)}\\  
parameters & 0 & 0.25 & 0.5 & 0.65 & 0.75 & 0.85 & 1 \\ \hline
Space group &$P2_1/n$&$P2_1/n$&$P2_1/n$&$P2_1/n$&$P2_1/n$&$P2_1/n$&$P2_1/n$\\
$a$ ({\AA})& 5.5820&5.5787&5.5713&5.5648&5.5612&5.5605&5.5566\\
$b$ ({\AA})& 5.6766&5.6419&5.6030&5.5991&5.5947&5.5885&5.5717\\
$c$ ({\AA})&7.9032&7.8974&7.8778&7.8728&7.8640&7.8622&7.8569\\
$\beta$ ($^\circ$)&89.98&90.074&89.99&90.017&89.972&89.991&89.995\\
$V$ ({\AA}$^3$)&250.50&248.57&246.47&245.43&244.61&244.10&243.00\\
$\chi^2$ &1.95&1.49&1.73&1.55&1.5&1.36&1.22\\
$R_{p}$ (\%) &12.1 &10.1 &27.5 &28.3&14.3&19.1&5.3\\
$R_{wp}$ (\%) &9.43 &9.2 &28.4 &18.1&11.8&12.6&6.69\\
$R_{exp}$ (\%) &6.75 &6.5 &21.6 &14.53&8.61&10.85&6.1\\
\hline
\hline
\end{tabular}
\caption{Refined structural parameters determined for La$_2$CoIr$_{1-x}$Ti$_x$O$_6$ under ambient conditions, providing detailed insights into the material's configuration at room temperature. There is a systematic change in the lattice parameters as a function of doping concentration.}
\label{tab:table 1}
\end{table*}

\subsection{Crystal structure}

\begin{figure*}[h!bt]
\centering
\includegraphics[width=2.0\columnwidth]{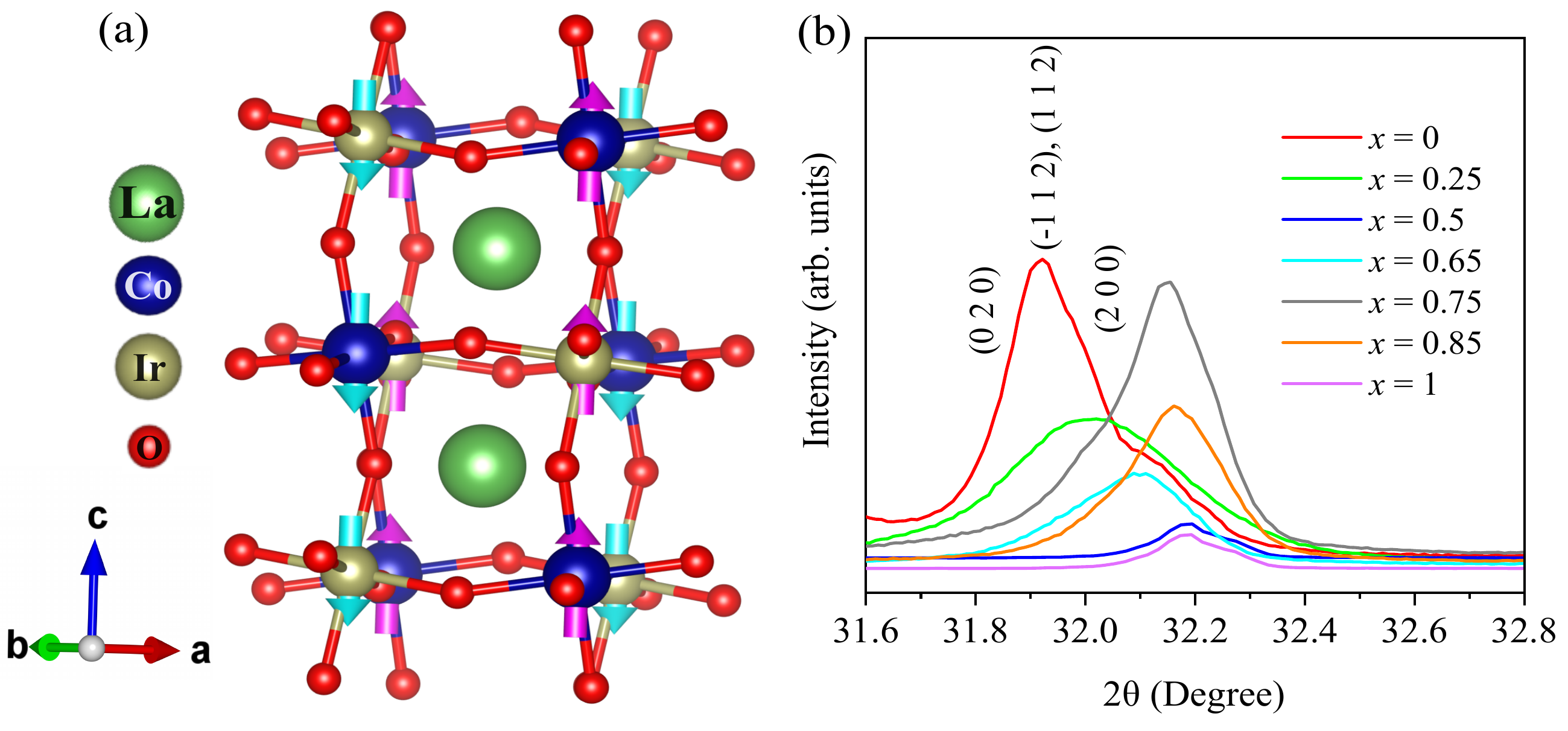}
\caption{\label{fig:XRD_shift_LCIO.pdf} (a) Crystal and magnetic structure of LCIO, where the structure is formed by corner-sharing CoO$_6$ and IrO$_6$ octahedra with La ions at the A sites. The Co$^{2+}$ and Ir$^{4+}$ spins are antiferromagnetically aligned within their sublattices and also with respect to each other, resulting in an overall AFM ground state. This magnetic structure is employed for the DFT calculations, and (b) the intensity of various Bragg's positions with the angle of La$_2$CoIr$_{1-x}$Ti$_x$O$_6$. The phenomenon of peak shifting in the XRD patterns is observed across all concentrations, revealing significant insights into the structural variations within the material. As the concentrations vary, the diffraction peaks gradually shift towards higher angles, indicating alterations in lattice parameters and atomic arrangements. In this figure, the ($hkl$) values are mentioned only for the parent compound (LCIO).}
\end{figure*} 

X-ray powder diffraction confirms that the parent and the entire doping series crystallize in a single phase with monoclinic space group $P2_1/n$ (No.~14), shown in Fig.~\ref{fig:xrd_plot.pdf}. The structural details for the parent compounds are in agreement with earlier reports~\cite{Narayanan2010,Nandi2024} and the quality of the Rietveld refinement is satisfactory ($\chi^{2}$: {1.22--1.95}) for all the samples. The unit cell volume, calculated from the structural refinement, of La$_2$CoIr$_{1-x}$Ti$_x$O$_6$ compounds decreases almost linearly with the doping level. Such linear Vegard’s law~\cite{Vegard1991} behavior and lack of secondary phases indicate that the Ti$^{4+}$ ion is being incorporated in the ordered perovskite structure at the intended levels. The crystal and magnetic structure of LCIO (one of the parent compounds) is shown in Fig. ~\ref{fig:XRD_shift_LCIO.pdf}(a) where the structure is formed by corner-sharing CoO$_6$ and IrO$_6$ octahedra. The Co$^{2+}$ and Ir$^{4+}$ spins are antiferromagnetically aligned within their sublattices and also with respect to each other, resulting in an overall AFM ground state. This magnetic structure is employed for the DFT calculations in section~\ref{sectionIII}. Also, the illustrated XRD pattern in Fig.~\ref{fig:XRD_shift_LCIO.pdf}(b) shows a rightward shift in peak positions as the Ti content increases, as the ionic radius of Ti$^{4+}$ is small compared to Ir$^{4+}$. There is an anomalous trend for $x$ = 0.5 may be due to the overlapping nature of the reflections or local structural distortion. Indeed, doping Ti$^{4+}$ at the Ir$^{4+}$ site introduced chemical pressure in the system, which is responsible for the decrement of the unit cell volume as $x$ increases. The lattice parameters for all the compounds are given in table~\ref{tab:table 1}.  





\subsection{Magnetization measurements}

\begin{figure*}[h!bt]
\centering
\includegraphics[width=1.95\columnwidth]{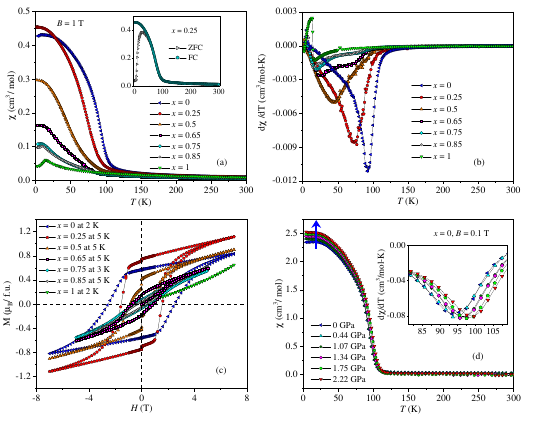}
\caption{\label{fig:Fig2_exp.pdf} (a) Temperature dependent magnetic susceptibilities of La$_2$CoIr$_{1-x}$Ti$_x$O$_6$, data taken upon FC with field of 1 T, (b) temperature derivative of magnetic susceptibility where the dip in the curves showing the transition temperature, (c) field dependence of magnetization for La$_2$CoIr$_{1-x}$Ti$_x$O$_6$ well below the magnetic ordering temperature, (d) temperature dependent magnetic susceptibility of LCIO at various constant hydrostatic pressure values for FC in the presence of field 0.1 T, and inset shows the derivative of magnetic susceptibility with temperature where the peak for different pressure shows the transition temperature. Note that the sample used for magnetic measurement under pressure (see Fig.~\ref{fig:Fig2_exp.pdf}(d)) and the sample used for other magnetic measurements (see Fig.~\ref{fig:Fig2_exp.pdf}(a) and (c)) are taken from two batches, indicating a sample dependency.)}
\end{figure*}

\begin{figure*}[h!bt]
\centering
\includegraphics[width=2\columnwidth]{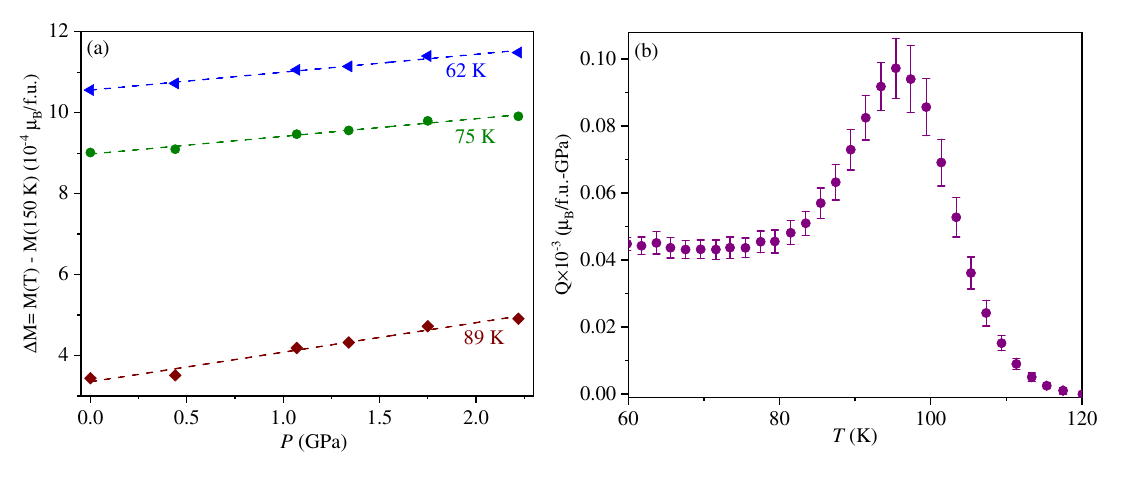}
\caption{\label{fig:PZM.pdf} (a) The relative change in the magnetization from the paramagnetic phase calculated using the relation $\Delta M$ = $M$($T$) - $M$(150 K), as a function of pressure at 62, 75, and 89 K. The dotted line indicates the linear fit using the function $\Delta M = M_0 + QP$, where $M_0$ stands for the spontaneous magnetization, $Q$ the powder-averaged piezomagnetic tensor, and $P$ the applied hydrostatic pressure. (b) Temperature dependence of the powder-averaged piezomagnetic tensor ($Q$).}
\end{figure*}

\begin{table*}
\begin{ruledtabular}
\begin{tabular}{cccccc}
$x$ & Transition temperature (K) & $\mu_{\mathrm{eff}}$ ($\mu_{\mathrm{B}}/ f.u.$) & $\theta_{\mathrm{CW}}$(K) & Bifurcation temperature (K)\\ \hline 
0 & 92.0 & 4.7 & -7.4~\cite{Narayanan2010} & 91.2\\
0.25 & 72.6 & 4.64 & -15.0 & 76.0\\ 
0.5 & 44.5 & 4.5 & -19.3 & 72.6\\ 
0.65 & 27.0 & 3.7  & -19.0 & 61.0\\
0.75 & 21.0  & 4.08 & -19.04 & 56.0\\
0.85 & 15.3 & 4.48 & -21.9 & 46.5\\
1 & 14.6 & 5.05 & -50.0~\cite{Nandi2024} & $-$\\
\end{tabular}
\caption{\label{tab:table 2} Curie-Weiss temperature, effective magnetic moment, magnetic ordering temperature, and bifurcation temperature for La$_2$CoIr$_{1-x}$Ti$_x$O$_6$. All the parameters show systematic variation with doping concentration, except the effective magnetic moment could be due to the deviation from the ideal 180$^\circ$ angle between Co-O-Ir sublattices and the reduction of the number of exchange pathways between Co and Ir as the Ti concentration increases at the Ir site.} 
\end{ruledtabular}
\end{table*}

Fig.~\ref{fig:Fig2_exp.pdf}(a) shows the temperature variation of magnetization of La$_2$CoIr$_{1-x}$Ti$_x$O$_6$ for all the compositions, measured at 1 T. FC data are shown for clarity, however, the measurements are done for both ZFC and FC conditions. For $x = 0$ (LCIO) shows a ferromagnetic (FM)-like transition at $T_\mathrm{C} =$ 92 K whereas for $x = 1$ (LCTO) it is an antiferromagnet (AFM) below $T_\mathrm{N} =$ 14.6 K, consistent with the literature~\cite{Narayanan2010,Nandi2024}. The values the $T_\mathrm{C}$ for all the samples ($x$ = 0, 0.25, 0.5, 0.65, 0.75, and 0.85) are obtained from the minimum of the temperature-derivative of magnetization in the $M~vs.~T$ curves, shown in Fig.~\ref{fig:Fig2_exp.pdf}(b). The FM-like transition temperature for $x = 0$ decreases gradually as the doping concentration increases and finally it transforms to an AFM state for $x = 1$. The bifurcation between ZFC and FC measurements, typical for a ferromagnet, occurs for all the concentrations before the system transforms to an AFM, shown in the inset of Fig.~\ref{fig:Fig2_exp.pdf}(a). There is nonmonotonic behavior of the maximum magnetization, $i.e.$, the values of the magnetization below $T \lesssim 25 K$ for $x = 0.25$ are slightly higher as compared to $x = 0$. However, it decreases as $x$ increases further. The Curie-Weiss fitting in the paramagnetic region for all the concentrations is done for 1/$\chi$ $vs.$ $T$. The observed magnetic ordering temperatures, effective magnetic moments, Curie-Weiss temperatures, and the bifurcation temperatures for all the concentrations are shown in table~\ref{tab:table 2}. All the parameters show a systematic trend with doping, except the effective magnetic moment that could be due to the deviation from the ideal 180$^\circ$ angle between Co-O-Ir sublattices and reduction in magnetic exchange pathways between Co and Ir. The Curie-Weiss temperature is negative for all the samples, however, it increases from $x = 0$ to $x = 1$, indicating the influence of ferromagnetic component except $x = 1$. This is supported by the vanishing of the bifurcation temperature between the ZFC and FC measurements and $M~vs.~H$ loop for $x = 1$.  

Isothermal $M~vs.~H$ measurements are performed at 2 and 5 K in the presence of external magnetic fields up to 7 T, shown in Fig.~\ref{fig:Fig2_exp.pdf}(c). The loop size decreases with the increase of Ti-concentration at the Ir-site. The absence of magnetic saturation for the series may also be related to the Co/Ir cationic disorder present in the whole series, which may lead to phase competition and magnetic frustration due to Co–Co and Ir–Ir interactions~\cite{Song2017}. The step-like feature in the $M~vs.~H$ curve is observed for all the doping series with partial Ti content except $x = 0$, which could be due to the sample dependency. It can be also observed from Fig.~\ref{fig:Fig2_exp.pdf}(a) and (d) where the measurements have been done on differently prepared LCIO samples. The presence of two magnetic sublattices (Co and Ir) brings out the step in the $M~vs.~H$ curve~\cite{Narayanan2010,Vogl2018}. Narayanan \textit{et~al.} attribute this step to a field-dependent change of the canting angle of the noncollinear magnetism in La$_2$CoIrO$_6$. The vanishing loop size and an almost linear field-dependent magnetization for $x = 1$ indicate that the system transforms to an antiferromagnet. We tried to identify the critical concentration, however, till $x = 0.85$ the system still shows a weak ferromagnetic behavior. 

Since chemical pressure strongly influences the magnetic properties, we applied hydrostatic pressure on LCIO to explore possible comparisons. With the increment of hydrostatic pressure, the transition temperature increases linearly from 92 K to 100 K at ambient pressure and $P$ = 2.22 GPa, respectively, shown in Fig.~\ref{fig:Fig2_exp.pdf}(d), with ${dT_\textrm{N}}/ {dP} \approx $ (3.6$\pm$0.5) K/ GPa. In addition, there is also a linear increment of the absolute magnetic moment at low temperatures as pressure increases. As the ground state of the system is noncollinear AFM, the application of pressure could help the spins to align more along the field, giving rise to a higher transition temperature and large magnetic moment in the ordered state. The trend of the transition temperature for chemical and physical pressure seems opposite which could be due to the effect of inhomogeneous pressure effect due to chemical doping~\cite{Fernandes1991}. The observed magnetization was subtracted from the value at 150 K in the paramagnetic state, $\Delta M$ = $M(T) - M(150$ K), shows a linear increase in magnetization value with the increase of pressure for different temperature values (see Fig.~\ref{fig:PZM.pdf}(a)) indicating the presence of piezomagnetic effect~\cite{Nanjo2025}. The magnetization data is fitted with the equation, $\Delta M = M_0 + QP$, and the obtained value of the spontaneous magnetization and the powder-averaged piezomagnetic tensor are $M_0$ = (8.97$\pm$0.03)$\times$10$^{-4}$ $\mu_\textrm{B}$/f.u. and $Q$ = (4.36$\pm$0.29)$\times$10$^{-5}$ $\mu_\textrm{B}$/f.u.-GPa, respectively, at 75 K. The powder-averaged piezomagnetic tensor as a function of temperature is depicted in Fig.~\ref{fig:PZM.pdf}(b). We emphasize that these results indicate a possible piezomagnetic response in LCIO, though a more detailed study would be required to establish this effect in other concentrations.

\subsection{Thermal expansion measurements}

\begin{figure}[h!bt]
\centering
\includegraphics[width=1.0\columnwidth]{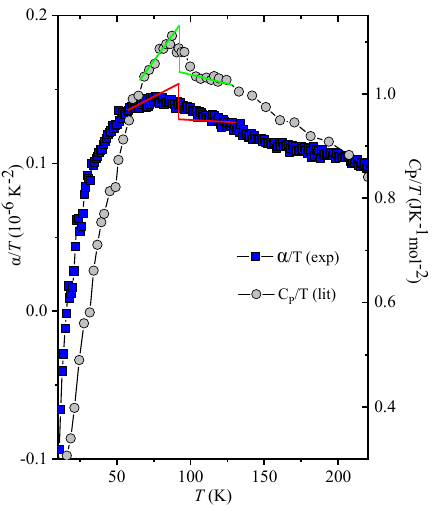}
\caption{\label{fig:TE.pdf} Temperature dependence of $\alpha/T$ and $C_\textrm{P}/T$~\cite{Narayanan2010_thesis} of LCIO. A broad transition is visible around the magnetic phase transition, indicating the coupling of lattice degrees of freedom. Entropy conserved equal-area construction for the thermal expansion (red solid line) and heat capacity (green solid line), respectively) was employed in order to determine the transition temperature.}
\end{figure}

Dilatometry serves as a valuable thermodynamic probe, providing insights into material properties through precise measurements of dimensional changes under varying temperatures. The linear thermal expansion coefficient, $\alpha$, defined as $d[\Delta L(T)/L_{0}]/dT$, is determined by analyzing the differential length change over temperature intervals of 0.5 K. Measurements are performed on a pressed pellet with a length of 1.9 mm, utilizing a temperature sweep rate of $+0.3$ K/minute. As temperature decreases, $\alpha/T$ increases almost linearly with temperature and then shows a subtle yet distinct broad hump around 92 K, indicating the coupling of lattice degrees of freedom with the spin degrees of freedom at the magnetic phase transition, illustrated in Fig.~\ref{fig:TE.pdf}. Notably, a similar transition has been reported through heat capacity measurements, as documented in~\cite{Narayanan2010_thesis}, indicating that the system adheres to the conventional Gr\"{u}neisen relation, as outlined in~\cite{Barron1980}. Entropy conserved an equal-area construction~\cite{Manna2014_VBS,Manna2014} is employed to determine the transition temperatures for both thermal expansion and specific heat measurements, typical for a second-order phase transition. According to the Ehrenfest relation~\cite{Manna2014_VBS,Manna2018}, $dT_\textrm{N}/dP = V_\textrm{mol}\times~T_\textrm{N}\times \Delta\beta/ \Delta C_\textrm{p}$, where $\Delta\beta$ and $\Delta C_\textrm{p}$ are the change in volume thermal expansion, and specific heat upon the transition, respectively. By taking molar volume = 7.54$\times$10$^{-9}$ m$^3$ mol$^{-1}$, the estimated pressure dependence of the transition temperature for LCIO in the zero pressure limit is (5.9$\pm$0.5) K/ GPa which is in well agreement with our estimated value from the magnetization measurements under pressure data.




\section{Electronic band structure calculations} \label{sectionIII}

\begin{figure}[hbt!]
\centering
\includegraphics[width=1.1\columnwidth]{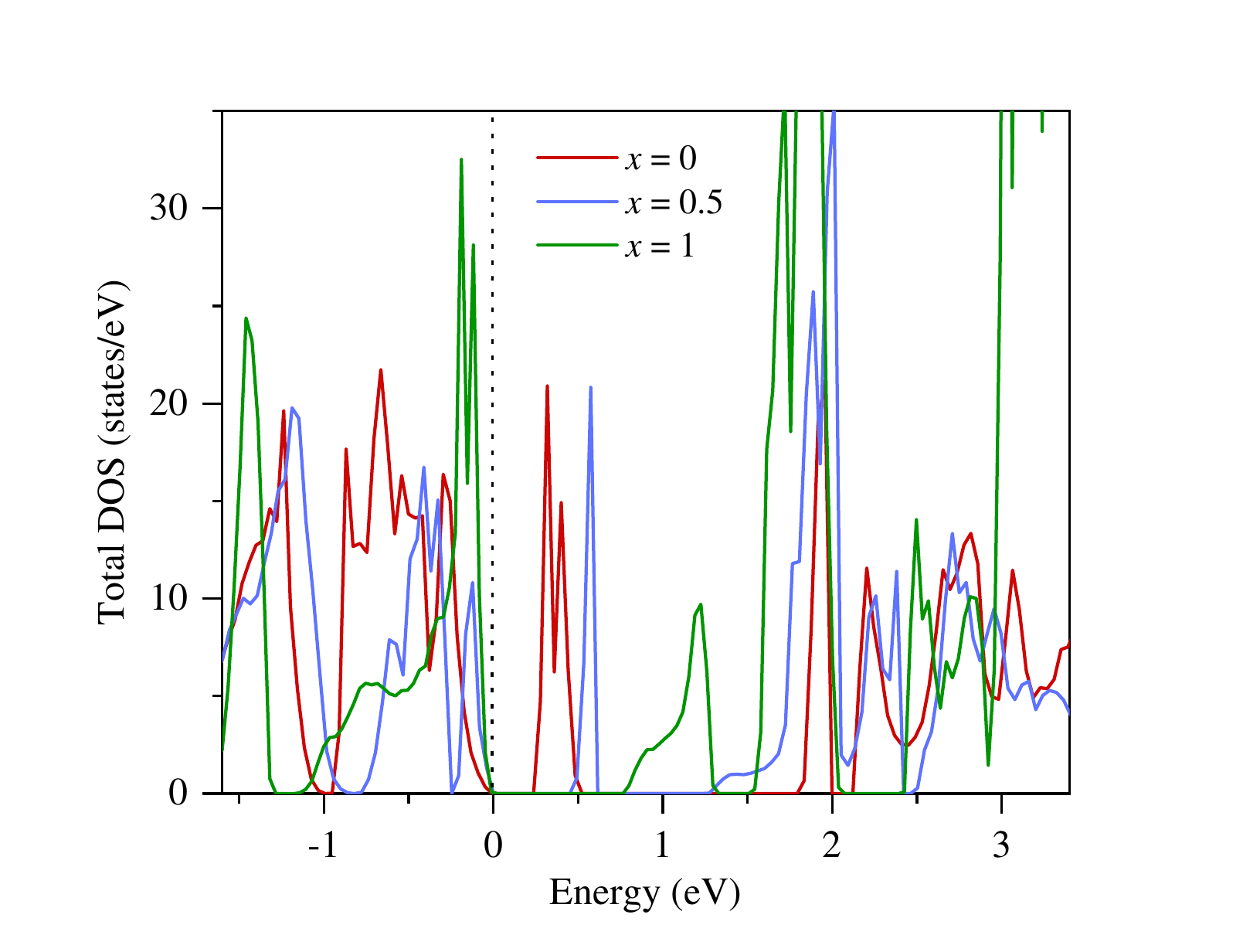}
\caption{\label{fig:LCIO_structure.pdf} Total density of states (TDOS) for $x = 0$, $x = 0.5$, and $x = 1$ calculated using the DFT-GGA combined with SOC and correlation $U$. Note that the gap value for $x$ = 1 is slightly reduced after adding SOC to GGA + $U$ calculations, expected for 3$d$ oxides where the effect of SOC is much less.
Quantization axes are taken in the direction [001] for SOC calculations. In all cases, the Fermi level ($E_\textrm{F}$) is set to zero.} 
\end{figure}

\begin{figure*}[h!bt]
\centering
\includegraphics[width=2\columnwidth]{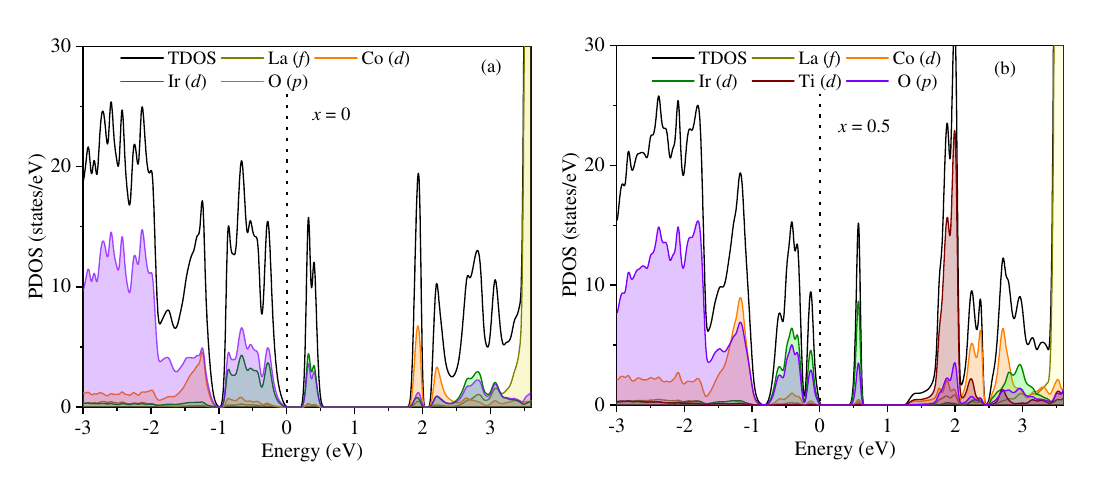}
\caption{\label{LCIO_totalpdos} The panels (a) and (b) show the total density of states (TDOS) and atom-projected partial density of states (PDOS) for LCIO ($x = 0$) and La$_2$CoIr$_{0.5}$Ti$_{0.5}$O$_6$ ($x = 0.5$) respectively. Here, GGA + SOC + $U$ is included in the DFT calculations. The black solid line represents the TDOS, while the PDOS for respective elements (Ir, La, Co, Ti, and O) are shown by the colored lines as indicated in panels (a) and (b). The $U$  values for Co and Ir are taken as 5 and 2 eV, respectively.}
\end{figure*} 

\begin{figure*}[h!bt]
\centering
\includegraphics[width=2.15\columnwidth]{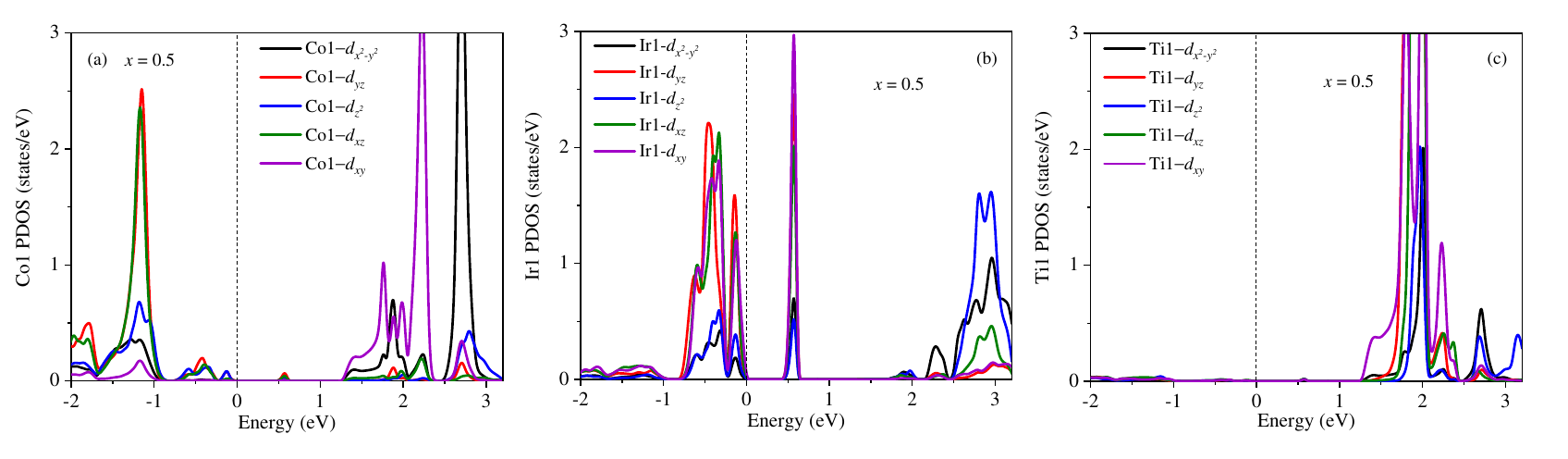}
\caption{\label{LCITO_orb} The panels (a), (b), and (c) show the orbital-projected PDOS for La$_2$CoIr$_{0.5}$Ti$_{0.5}$O$_6$ ($x = 0.5$) for Co, Ir, and Ti-$d$ orbitals, respectively. This calculation has been done under GGA + SOC + $U$ ($U_\mathrm{Co}$ = 5 eV and $U_\mathrm{Ir}$ = 2 eV) where the quantization axis is along [001] direction.}
\end{figure*} 

The electronic and magnetic properties of both the parent compounds including 50\% doping, $i.e.$, $x$ = 0, 0.5 and 1 at ambient conditions have been explored within \textit{ab~initio} spin-polarized electronic structure calculations using projector augmented wave (PAW) scheme~\cite{Blochl1994,Kresse1999} in the plane wave pseudopotential implementation of DFT code, Vienna \textit{Ab~initio} Simulation Package (VASP)~\cite{Kresse1996}. Perdew-Burke-Ernzerhof (PBE)~\cite{Perdew1996} functional within generalized gradient approximation (GGA) was adopted for the exchange correlation potential. A $k$-mesh of $6\times6\times4$ ($\Gamma$-centered) was used for the momentum space integration over a full Brillouin zone, while a plane wave cutoff energy of 500 eV was used for the plane wave basis expansion in all calculations. To account for the electron-electron correlation effects on Co-3$d$ electrons and the impact of SOC originating from Ir-5$d$, a sequence of schemes: GGA, GGA + SOC, and GGA + SOC + $U$, was applied within the formalism introduced by Kune\v{s} \textit{et~al.}~\cite{Kunes2001}. Here, we varied $U$ value between 3 to 5 eV on Co-$d$ (3$d^7$) states while 2 to 3 eV on Ir (5$d^5$), similar to the reported literature~\cite{Narayanan2010,Ganguly2020}. As Ti (3$d^0$) has empty $d$-shell configuration, we kept the $U$ value for Ti as zero. Indeed, for LCTO, we performed the calculations with $U$ = 1 and 2 eV for Ti, however, no significant change in the electronic structure was observed compared to $U$ = 0. The internal positions of all atoms were meticulously relaxed until the forces diminished below 10$^{-3}$ eV/{\AA}, with an energy convergence cutoff set at 10$^{-7}$ eV. 

A systematic investigation of different possible collinear magnetic configurations of LCIO is explored: (a) Co$^{2+}$ and Ir$^{4+}$ sublattices are ferromagnetically coupled within their respective sublattices and with each other, resulting into an overall FM interaction, (b) FM coupling within Co$^{2+}$ sublattice with all the spins aligned up and similarly within the Ir$^{4+}$ sublattice with all spins aligned down, leading to an overall AFM interaction, (c) AFM coupling within both Co$^{2+}$ and Ir$^{4+}$ sublattices and with each other, resulting into an overall AFM interaction. The ground state magnetic configuration, which corresponds to the minimum energy, is represented in Fig.~\ref{fig:XRD_shift_LCIO.pdf}(a). This unveils the magnetic structure where Co$^{2+}$ and Ir$^{4+}$ are antiferromagnetically aligned within their sublattice and with each other showing commensurate arrangement also described by the propagation vector $k$ = (0, 0, 0)~\cite{Narayanan2010}. This ordered antiferromagnetic state manifests itself at temperatures marginally below 92 K, a phenomenon that has been scrutinized as ferromagnetic-like in prior experimental studies~\cite{Narayanan2010}. The system is characterized by distinct magnetic B-3$d$ sites housing Co$^{2+}$ ions and B$^\prime$-3$d$ sites hosting Ir$^{4+}$ ions. Both GGA and GGA + SOC predict metallic behavior (not shown), unlike the insulating behavior observed experimentally. However, adding electronic correlation $U$, $i.e.$, GGA + SOC + $U$ opens up a band gap of 0.28 eV as shown in Fig.~\ref{fig:LCIO_structure.pdf}, which is in good agreement with the literature~\cite{Narayanan2010}. The values assigned for $U$ and $J$ (Hund's exchange coupling) are $U_\mathrm{Co} = 5$ eV, $J_\mathrm{Co} = 1.03$ eV, $U_\mathrm{Ir} = 2$ eV, and $J_\mathrm{Ir} = 0.94$ eV and these values remained unchanged for $x = 0$ and $x = 0.5$ except $x = 1$. For $x$ = 1, the $U$ value of 2 eV is assigned to the Co. The effect of SOC is significant for the magnetic moment of Ir as it is increased from 0.024 $\mu_\textrm{B}$/Ir (GGA + $U$) to 0.315 $\mu_\textrm{B}$/Ir (GGA + SOC + $U$) whereas the moment of Co remains the same for the GGA + $U$ and GGA + SOC + $U$ calculations. 

\begin{table}[ht]
\begin{ruledtabular}
\begin{tabular}{cccc} 
$x$ & Magnetic moment ($\mu_\textrm{B}$) & Volume (\AA$^3$) & Band gap $\Delta_{g}$ (eV)
\\ [0.3ex] 
\hline 
0 & $\mu_\textrm{Co}$ = 2.686, $\mu_\textrm{Ir}$ = 0.315 & 250.7 & 0.28~\cite{Narayanan2010}\\
0.5 & $\mu_\textrm{Co}$ = 2.67, $\mu_\textrm{Ir}$ = 0.126 & 245.4 & 0.44\\
1 & $\mu_\textrm{Co}$ = 2.6 & 243.5 & 1.01~\cite{Nandi2024}\\
\end{tabular}
\caption{\label{tab:table1} The table displays the variation of magnetic moment, unit cell volume, and band gap of La$_2$CoIr$_{1-x}$Ti$_x$O$_6$ for $x = 0, 0.5$ and 1 providing valuable insights into the material's magnetic, structural, and electronic properties across different compositions. The values of the magnetic moment, volume, and band gap for $x$ = 0 and 1 are in close agreement with the literature~\cite{Narayanan2010,Nandi2024}.}
\label{table:nonlin} 
\end{ruledtabular}
\end{table}

Now we focus on the doped sample, especially for $x = 0.5$ to study the effect of magnetic and transport properties with 50\% dilution with the nonmagnetic Ti at the Ir-site. There is a contraction of the unit cell volume as compared to the $x = 0$ due to the effect of chemical pressure. However, the structural symmetry remains unchanged ($P2_1/n$). Doping influenced the electronic properties of the material, $viz.$, the band gap increases to 0.44 eV. For $x$ = 0.5, we have varied the $U$ value of Co from 2--5 eV ($U_\mathrm{Ir} = 2$ eV fixed), but the gap was found to be almost unchanged for $U_\mathrm{Co} \geq 3$ eV. Fig.~\ref{fig:LCIO_structure.pdf} shows the total DOS for the doped compounds along with the parent compounds LCIO and LCTO. The magnetic moment of Ir decreases significantly due to Ti doping, highlighting a impact on the magnetic properties of LCIO due to a nonmagnetic chemical doping. Indeed, the magnetic moment increased from 0.014 $\mu_\textrm{B}$/Ir (GGA + $U$) to 0.126 $\mu_\textrm{B}$/Ir (GGA + SOC + $U$). The moment on Ir decreases by more than 50\% with the nonmagnetic Ti-doping, indirectly showing the effect of Ti on moments via SOC, given in table~\ref{tab:table1}. This dilution in the magnetic moment at the Ir site may also be attributed to the weakening of the nearest neighbor Co-O-Ir superexchange interaction and the disruption of the second nearest neighbor Co-O-Ir-O-Co pathways, which are replaced by the nonmagnetic Co-O-Ti path connection, in which Ti atoms creates an empty states about 2 eV away from the Fermi level (see Fig.~\ref{LCIO_totalpdos}(b)).

Indeed Fig.~\ref{LCIO_totalpdos} shows the atom-projected PDOS for $x$ = 0 and 0.5 compounds, where a noticeable shift in the conduction band states is observed for $x$ = 0.5, resulting in an increased energy gap at the Fermi level. Additionally, the energy range between $-3$ eV and $-1$ eV shows considerable hybridization between O-$p$ and Co-$d$ states, with a minor contribution from Ir-$d$ orbitals. The contribution of the Ti-$d$ states in $x$ = 0.5 to the conduction band is quite pronounced as compared to the narrow Ir-$d$ states in the energy range of 1.5 to 2.5 eV, highlighting the significant impact of Ti doping on the system.

Subsequently, we analyzed the orbital-projected PDOS for sublattices Co1 and Ir1-$d$ under GGA + SOC + $U$ scheme for $U_\mathrm{{Co}}$ = 5 eV and $U_\mathrm{{Ir}}$ = 2 eV values, shown in Fig.~\ref{LCITO_orb} for $x$ = 0.5. For the $x$ = 0.5, there is a significant increase in the band gap of 0.44 eV, which is attributed to the modified electronic structure due to Ti doping. The band gap opens as direct consequence of the fewer states of Ti1-$d$ and Co1-$d$ available in the conduction band near the Fermi level.

The $t_{2g}$ and $e_g$ states are available for Co1-$d$ in the energy range of 1.25 to 2.3 eV, and then from 2.5 to 3 eV, respectively, as shown in Fig.~\ref{LCITO_orb}(a). The empty but slightly narrow states (Fig.~\ref{LCITO_orb}(b)) are again visible for Ir1-$d$ state near the Fermi level in the energy range from 0.4 to 0.6 eV. Note that Ti1-$d$ plays no role near the Fermi level, see Fig.~\ref{LCITO_orb}(c). As the magnetic dilution at the Ir-site gives a significant effect in the magnetic and transport properties, it underlines the importance of SOC in the system.   

Identifying the critical Ti concentration may provide useful information to understand this transition, a scope of future studies. The presence of two distinct magnetic sublattices Co and Ir with canted spin configurations, and the observed suppression of the Ir sublattice upon Ti doping, suggest a complex evolution of magnetic structure. Notably, increased hydrostatic pressure in LCIO compounds appears to enhance spin alignment, indicating that neutron diffraction studies under pressure could provide valuable insights into the nature of magnetic ordering across the doping series.

\section{Summary and conclusions} \label{sectionIV}

In this comprehensive exploration, La$_2$CoIr$_{1-x}$Ti$_x$O$_6$ was meticulously synthesized and characterized structurally, magnetically, and thermodynamically across a range of compositions ($x$ = 0, 0.25, 0.5, 0.65, 0.75, 0.85, and 1) using the solid-state method, complemented by diligent first-principle calculations for $x$ = 0, 0.5 and 1. The Ti$^{4+}$ doping at the Ir$^{4+}$ site gradually decreases the unit cell volume, thus indicating the influence of chemical pressure. This effects the magnetic properties, $viz.$, there is a gradual suppression of the FM-like transition from $x$ = 0 ($T_\textrm{c}$ = 92 K) which leads to an AFM transition at $x$ = 1 ($T_\textrm{N}$ = 14.6 K). Thermal expansion measurements on $x$ = 0 show the magnetic transition is coupled with the lattice degrees of freedom, and the initial pressure dependency of the transition temperature, obtained from the Ehrenfest relation, is consistent with the magnetization under hydrostatic pressure measurements. The piezomagnetic effect has been observed for $x$ = 0, which is characterized by a linear relationship between applied stress and induced magnetization. The transport properties are analyzed using the first-principle calculations, where the band gap widens on increasing the Ti$^{4+}$ concentration as follows: 0.28 eV for $x$ = 0, 0.44 eV for $x$ = 0.5, and 1.01 eV for $x$ = 1. Also, the Ir$^{4+}$ magnetic moment is reduced more than 50\% for $x$ = 0.5. DFT studies on other concentrations both at ambient and hydrostatic pressure conditions may provide a relation between the chemical and physical pressure effects in this system. These theoretical analyses offer a pathway to elucidate the underlying mechanisms governing observed behaviors, potentially unveiling new facets of magnetic and electronic properties in these complex materials. The robustness of the ferromagnetic state in the Ir-based double perovskite towards Ti substitution may underscore its potential for diverse applications.

\vspace{2mm}

\begin{acknowledgments}

RSM acknowledges the financial support from SERB for his Early Career Research Award with file number ECR/2018/000999/PMS. The authors would like to acknowledge IIT Tirupati for providing an experimental setup for sample preparation and high-performance computing (HPC) cluster for DFT calculations. SD acknowledges financial support from Technical Research Center, IACS and DST-SERB grant with file number CRG/2021/004334. 

\end{acknowledgments}

\end{document}